\documentclass[twocolumn,aip]{revtex4}

\usepackage{graphicx}
\usepackage{dcolumn}
\usepackage{bm}
\usepackage{color}
\usepackage{setspace}

\begin{document}


\title{Interface-dependent magnetotransport properties for thin Pt films on ferrimagnetic Y$_{3}$Fe$_{5}$O$_{12}$}

\author{Y. Shiomi$^{\, 1}$} 
\author{T. Ohtani$^{\, 1}$} \thanks{Y. Shiomi and T. Ohtani contributed equally to this work.}
\author{S. Iguchi$^{\, 1}$}
\author{T. Sasaki$^{\, 1}$}
\author{Z. Qiu$^{\, 2}$}
\author{H. Nakayama$^{\, 1,3}$}
\author{K. Uchida$^{\, 1,4}$}
\author{E. Saitoh$^{\, 1,2,5,6}$}
\affiliation{$^{1}$
Institute for Materials Research, Tohoku University, Sendai 980-8577, Japan }
\affiliation{$^{2}$
WPI Advanced Institute for Materials Research, Tohoku University, Sendai 980-8577, Japan
}
\affiliation{$^{3}$
Laboratory for Nanoelectronics and Spintronics, Research Institute of Electrical Communication, Tohoku University, Sendai 980-8577, Japan
}
\affiliation{$^{4}$
PRESTO, Japan Science and Technology Agency, Saitama 332-0012, Japan
}
\affiliation{$^{5}$
CREST, Japan Science and Technology Agency, Tokyo 102-0076, Japan
}
\affiliation{$^{6}$
Advanced Science Research Center, Japan Atomic Energy Agency, Tokai 319-1195, Japan
}

\date{\today}

\begin{abstract}
We have studied magnetoresistance and Hall effects for $1.8$-nm-thick Pt films grown on a ferrimagnetic insulator Y$_{3}$Fe$_{5}$O$_{12}$ in a wide temperature ($0.46$-$300$ K) and magnetic-field ($-15$-$15$ T) region. In the low-temperature regime where quantum corrections to conductivity are observed, weak antilocalization behavior observed in Pt films is critically suppressed when the film is attached to Y$_{3}$Fe$_{5}$O$_{12}$. Hall resistance in the Pt film is also affected by Y$_{3}$Fe$_{5}$O$_{12}$, and it exhibits logarithmic temperature dependence in a broad temperature range. The magnetotransport properties in the high-field range are significantly influenced by the interface between Pt and Y$_{3}$Fe$_{5}$O$_{12}$. 
\end{abstract}

\maketitle

In the field of spintronics, a pure spin current, which is a flow of spin angular momentum without a net charge current, has attracted a great deal of attention in view of spin-current science and also of practical application \cite{spin-current}. For study on spin-current phenomena, Pt$\mid$Y$_{3}$Fe$_{5}$O$_{12}$ (Pt$\mid$YIG) bilayers have been used frequently as a typical system. YIG is a ferrimagnetic insulator with a large charge gap ($\sim2.7$ eV) and a high magnetic-transition temperature ($\sim 553$ K), which enables spin-current injection free from spin-polarized currents at room temperature. Injected pure spin currents are able to be detected electrically in Pt by means of the inverse spin Hall effect (ISHE) which is the conversion of an injected spin current into a transverse electric current due to the spin-orbit interaction. Since Pt has strong spin-orbit interaction, efficiency of the ISHE is as high as $1$-$10$ percent; hence, Pt has often been used as a spin-current detector. Using Pt$\mid$YIG systems, many experiments on spin-current injection and detection have been performed, {\it e.g.} spin pumping \cite{kajiwara} and the spin Seebeck effect \cite{uchida}. 
\par

Recently, an unconventional magnetoresistance (MR) effect was reported for Pt$\mid$YIG structures. Although Pt is a paramagnetic metal, MR in about $10$-nm-thick Pt films on YIG reflect the magnetization direction of YIG and anisotropic MR was observed in a low magnetic-field region ($\leq 0.2$ T) \cite{huang, nakayama}. This anisotropic MR was found to be caused mainly by a spin mixing effect at the interface between Pt and YIG \cite{nakayama}; concerted actions of the direct and inverse spin Hall effects generate an additional electric current and thus lead to resistance change affected by the magnetization direction in YIG. This magnetoresistance was named the spin-Hall magnetoresistance (SMR) \cite{nakayama} and this mechanism has been supported by following reports \cite{althammer, vlietstra, vlietstra-2, hahn, weiler-2, isasa, yang, geprags}.

\par

In the present paper, we discuss interface-dependent magnetotransport properties in Pt$\mid$YIG at low temperatures using very thin ($\sim 2$ nm) Pt films where the interface effect should be further pronounced owing to the reduced Pt volume. By conducting magnetotransport measurements in a wide temperature ($0.46$-$300$ K) and magnetic field ($-15$-$15$ T) region, we have shown that MR and Hall effects at high magnetic-fields in Pt$\mid$YIG exhibit totally different behavior from those in conventional paramagnetic metals. These unconventional magnetotransport properties are prominent at low temperatures and at high magnetic-fields, which are clearly irrelevant to magnetization change in YIG. 
\par

\begin{figure}[t]
\begin{center}
\includegraphics[width=8cm]{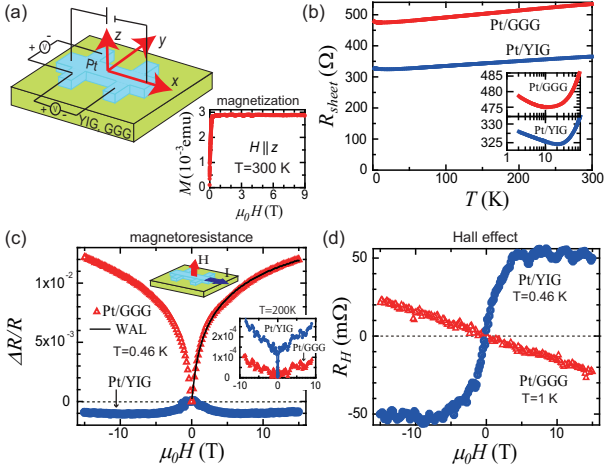}
\caption{(a) A schematic illustration of experimental setup. An electric current is applied along the $x$ axis. Longitudinal (Hall) resistance is calculated from the applied electric-current and voltage produced along the $x$ ($y$) axis. Magnetic field ($H$) dependence of magnetization ($M$) for a YIG film ($1.5$mm$\times 2$mm) at $300$ K in $H||z$. (b) Temperature ($T$) dependence of sheet resistance ($R_{sheet}$) for Pt$\mid$YIG and Pt$\mid$GGG. $R_{sheet}$ in a low-$T$ range is magnified in the inset to (b); the temperature scale is logarithmic. (c) Magnetic field ($H$) dependence of magnetoresistance [$\Delta R/R = \{R(H)-R(H=0)\}/R(H=0)$] in $H||z$ for Pt$\mid$YIG and Pt$\mid$GGG at $0.46$ K. The solid line is the fit to weak-anti localization (WAL). The inset shows results at $200$ K. (d) Magnetic field ($H$) dependence of Hall resistance for Pt$\mid$YIG at $0.46$ K and Pt$\mid$GGG at $1$ K.} 
\label{fig1}
\end{center}
\end{figure}

We measure magnetotransport properties of Pt thin films attached to (111) planes of YIG fims or paramagnetic Gd$_{3}$Ga$_{5}$O$_{12}$ (GGG) substrates. Here, Pt$\mid$GGG was used for control experiments, since GGG has the same crystal structure as YIG and is paramagnetic down to $0.46$ K. Micrometer-thick YIG films were grown on (111) GGG substrates by liquid phase epitaxy \cite{qiu}; the magnetization of YIG films is saturated for $\mu_{0}H \gtrsim 0.3$ T in a perpendicular magnetic field [Fig. \ref{fig1}(a)]. Before deposition of Pt, YIG films and GGG substrates were first cleaned in organic solvents inside an ultrasound bath, following surface treatment with H$_{2}$SO$_{4}$ and H$_{2}$O$_{2}$; this process is important to observe the interface-dependent magnetotransport phenomena in the present work. We then sputtered $1.8$-nm-thick Pt thin films with Hall-bar geometry on cleaned YIG or GGG surfaces in Ar pressure of $7.0$ mTorr. The magnetotransport measurements were performed as illustrated in Fig. \ref{fig1}(a). The measurements were carried out in superconducting magnets up to $\pm 15$ T in the temperature range between $0.46$ K and $2$ K as well as up to $\pm 9$ T in that from $2$ K to $300$ K.
\par

We show, in Fig. \ref{fig1}(b), the temperature ($T$) dependence of sheet resistance, $R_{\rm sheet}$, for Pt$\mid$YIG and Pt$\mid$GGG. The resistance for both the samples shows metallic $T$ dependence with the residual resistance $R_{\rm sheet} \sim 300$-$500$ ${\rm \Omega}$. Different magnitudes of $R_{\rm sheet}$ between Pt$\mid$YIG and Pt$\mid$GGG mainly originate from slightly different Pt-thicknesses which are inevitable in our sputtering system. As shown in the inset to Fig. \ref{fig1}(b), $R_{\rm sheet}$ shows a minimum around $20$ K and then increases with decreasing $T$ below $\sim 20$ K. This resistance rise is almost proportional to $\ln T$, indicating manifestation of weak (anti-)localization which is incipient of quantum corrections in disordered conductors.
\par

Figure \ref{fig1}(c) shows magnetic field ($H$) dependence of MR for Pt$\mid$YIG and Pt$\mid$GGG at $T=0.46$ K. Here, the magnitude of MR is defined as $\Delta R/R \equiv \{ R(H)-R(H=0)\}/R(H=0)$. For Pt$\mid$GGG, positive MR is observed and its magnitude is $\sim 1$\% at $15$ T. This positive MR in Pt$\mid$GGG is well explained by weak anti-localization (WAL) which appears in disordered conductors with strong spin-orbit interaction \cite{bergman, niimi}. By contrast, Pt thin films on YIG show a totally different MR effect from Pt$\mid$GGG, at $0.46$ K. The MR is negative and its magnitude is as small as $0.1$\%. This clear difference in MR between Pt$\mid$YIG and Pt$\mid$GGG is not observed at high temperatures; as shown in the inset to Fig. \ref{fig1}(c),  at $200$ K, while SMR reflecting the magnetization process of YIG is observed for Pt$\mid$YIG in a low-$H$ region ($<0.5$ T), MR effects in a high-$H$ regime ($>0.5$ T) are similar between Pt$\mid$YIG and Pt$\mid$GGG. These results clearly show that in a low-$T$ range where quantum corrections are observed, unconventional MR shows up in Pt$\mid$YIG at high magnetic-fields where the magnetization of YIG is fully aligned along the $H$ direction. 
\par

\begin{figure}[t]
\begin{center}
\includegraphics[width=8cm]{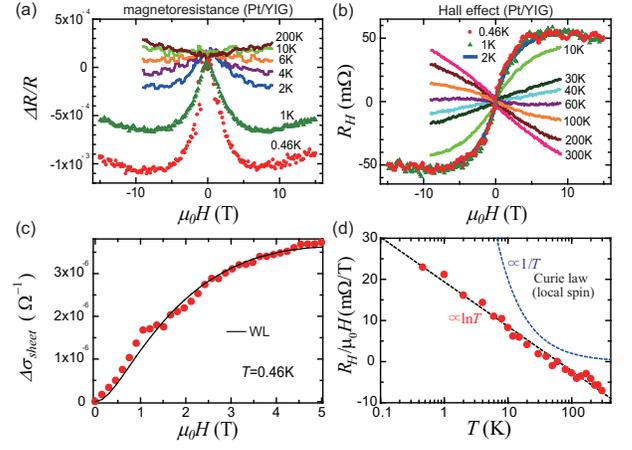}
\caption{Magnetic field ($H$) dependence of (a) magnetoresistance ($\Delta R/R$) in $H||z$ and (b) Hall resistance ($R_{H}$) at various temperatures between $0.46$ K and $300$ K. (c) $H$ dependence of magnetoconductance [$\Delta \sigma_{sheet} = 1/R_{sheet}(H) - 1/R_{sheet}(H=0)$] for Pt$\mid$YIG at $0.46$ K. Here, SMR contribution observed in a low-$H$ region is subtracted. The solid line is the fit to weak localization (WL). (d) Temperature ($T$) dependence of the Hall coefficient in the $T$ range between $0.46$ K and $300$ K. The Hall coefficient is calculated in the low-$H$ region $<1$ T. The temperature scale is logarithmic. The dotted lines are curves proportional to $\ln T$ and $1/T$, which are guides for the eyes.  } 
\label{fig2}
\end{center}
\end{figure}

\begin{figure}[t]
\begin{center}
\includegraphics[width=6cm]{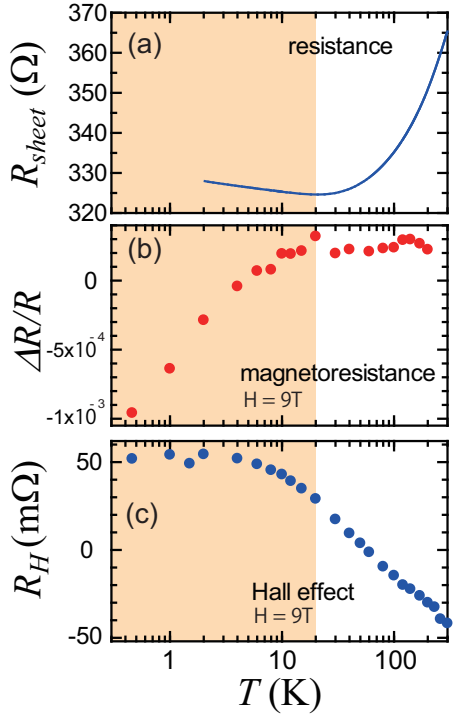}
\caption{Temperature ($T$) dependence of (a) sheet resistance ($R_{sheet}$), (b) magnetoresistance ($\Delta R/R$) at $\mu_{0}H=9$ T ($||z$), and (c) Hall resistance ($R_{H}$) at $\mu_{0}H=9$ T. Here, the temperature scale is logarithmic. Weak (anti-)localization regime is highlighted in yellow color. }
\label{fig3}
\end{center}
\end{figure}

\begin{figure}[t]
\begin{center}
\includegraphics[width=8cm]{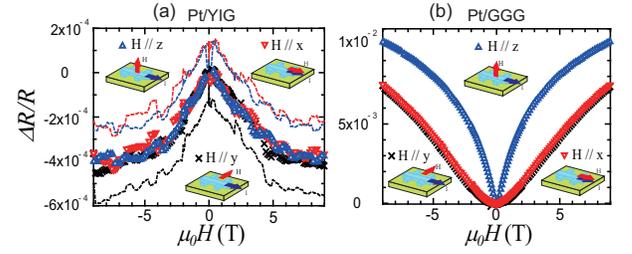}
\caption{Magnetic field ($H$) dependence of magnetoresistance ($\Delta R/R$) in three different magnetic-field directions ($H||x$, $H||y$, or $H||z$) for (a) Pt$\mid$YIG and (b) Pt$\mid$GGG. Measurement temperature is $2$ K. In (a), the dotted lines are raw data, while the symbols indicate the data in which SMR contributions observed in a low-$H$ region are subtracted. }
\label{fig4}
\end{center}
\end{figure}

In Fig. \ref{fig2}(a), we show MR in Pt$\mid$YIG at various temperatures. At $200$ K, positive MR showing quadratic $H$-dependence is observed in a high-$H$ region; this is characteristic of ordinary MR related with Lorentz force \cite{ziman}. As $T$ is decreased, MR hardly changes with $T$ down to $10$ K, but, below $10$ K, MR in a high-$H$ region shows a sign change from positive to negative and its magnitude abruptly increases, while SMR observed in a low $H$ region ($<0.5$ T) is almost independent of temperature even in this $T$ range [see also Figs. \ref{fig1}(c) and \ref{fig4}(a)]. We compare the $T$ range of MR enhancement and that of weak (anti-)localization regime determined from $T$-$R_{sheet}$ curve, in Figs. \ref{fig3}(a) and (b). As shown in Fig. \ref{fig3}(b), in the weak (anti-)localization regime (highlighted in yellow color in Fig. \ref{fig3}), negative MR at $9$ T is enhanced almost in proportion to $\ln T$, which signals weak localization (WL) behavior \cite{bergman}. On YIG, WAL in Pt is suppressed and WL appears in spite of the strong spin-orbit interaction in Pt.    
\par

In magnetic fields and under strong spin-orbit interaction, the quantum correction to the sheet conductance, $\Delta \sigma_{sheet} (H) \equiv 1/R_{sheet} (H) - 1/R_{sheet} (H=0)$, is given by \cite{hikami, maekawa, bergman} 
\begin{eqnarray}
\Delta \sigma_{sheet} (H)  &=& \frac{e^{2}}{2\pi^{2}\hbar} \Bigl[  \psi \Bigl( \frac{1}{2} +\frac{B_{1}}{\mu_{0}H}  \Bigr) -\frac{3}{2} \psi \Bigl( \frac{1}{2} +\frac{B_{2}}{\mu_{0}H}  \Bigr) \nonumber \\
& & + \frac{1}{2} \psi \Bigl( \frac{1}{2} +\frac{B_{3}}{\mu_{0}H}  \Bigr)  -\ln \Bigl( \frac{B_{1}\sqrt{B_{3}}}{B_{2}^{3/2}} \Bigr)  \Bigr],
\label{HK}
\end{eqnarray}
where $B_{1}=B_{\tau}+B_{SO}$, $B_{2}=(3/4)B_{SO}+B_{\phi}$, and $B_{3}=B_{\phi}$. Here, $B_{\tau}$, $B_{SO}$, and $B_{\phi}$ are effective magnetic fields for elastic, spin-orbit, and inelastic scatterings, respectively. Using eq.(1), we fit MR at $0.46$ K for Pt$\mid$GGG and Pt$\mid$YIG, as shown with solid lines in Figs. \ref{fig1}(c) and \ref{fig2}(c), respectively. The fitted curves almost reproduce the experimental results in both cases. The fitting parameters are $B_{\tau}=200$ T, $B_{SO}=20$ T, and $B_{\phi}=0.049$ T for Pt$\mid$GGG, while $B_{\tau}=200$ T, $B_{SO}=2.2$ T, and $B_{\phi}=0.31$ T for Pt$\mid$YIG. The decrease in $B_{SO}$ in Pt$\mid$YIG is explained by suppression of spin-flip scattering caused by spin-orbit interaction in the presence of magnetic spin-exchange interaction at the interface \cite{dugaev, sil, kurzweil}. We note that the change in $B_{SO}$ values between Pt$\mid$YIG and Pt$\mid$GGG also manifests itself in $R_{sheet}$ in the zero field; since the temperature of minimal sheet-resistance is proportional to $\ln(1/B_{SO})$ \cite{bergman}, that temperature is higher in Pt$\mid$YIG than Pt$\mid$GGG, as shown in the insets to Fig. \ref{fig1}(b). While $B_{SO}$ is smaller in Pt$\mid$YIG, $B_{\phi}$ is larger in Pt$\mid$YIG than Pt$\mid$GGG. A possible magnetic scattering around the interface \cite{takahashi, pd} may enhance the effective $B_{\phi}$ value in Pt$\mid$YIG.
\par

Such an interface effect also appears in the Hall effect. Figure \ref{fig1}(d) shows $H$ dependence of Hall resistance, $R_{H}$, at $0.46$ K for Pt$\mid$YIG and at $1$ K for Pt$\mid$GGG. $R_{H}$ for Pt$\mid$GGG shows linear dependence on $H$; this is the normal Hall effect induced by Lorentz force. In Pt$\mid$YIG, by contrast, $R_{H}$ shows clearly nonlinear $H$-dependence and its magnitude is much larger than that for Pt$\mid$GGG; with increasing $H$ from the zero field, $|R_{H}|$ increases dramatically and becomes almost saturated above $5$ T. This $H$ dependence of $R_{H}$ corresponds to neither the applied magnetic field nor the magnetization process in YIG. As shown in Fig. \ref{fig1}(d), the field value ($\sim 5$ T) where $R_{H}$ becomes almost saturated is much higher than the saturation field of YIG magnetization ($\sim 0.3$ T), which indicates that the internal magnetic field induced by YIG magnetization is not the origin of the nontrivial $H$ dependence of $R_{H}$.        
\par

A plot of Hall resistance ($R_{H}$) versus $H$ is shown at various temperatures between $0.46$ K and $300$ K in Fig. \ref{fig2}(b). While MR largely changes only at low temperatures below $10$ K, $R_{H}$ depends on $T$ even above $100$ K. $R_{H}$ in Pt$\mid$YIG significantly changes with $T$ and even shows a sign change around $60$ K. Since the normal Hall effect in paramagnetic metals is independent of $T$ as observed in Pt$\mid$GGG (not shown), this sign change suggests the presence of another contribution to the Hall effect other than the normal Hall effect in Pt$\mid$YIG: anomalous Hall effect \cite{ahe-review} or topological Hall effect \cite{the}.
\par

We found that the Hall coefficient defined as $R_{H}/(\mu_{0}H)$ in a low-$H$ region below $1$ T shows logarithmic $T$ dependence in all the $T$ region between $0.46$ K and $300$ K, as shown in Fig. \ref{fig2}(d). In very recent papers \cite{takahashi, pd}, the origin of similar nonlinear Hall resistance in Pt$\mid$YIG \cite{takahashi} and Pd$\mid$YIG \cite{pd} was attributed to the anomalous Hall effect assuming local {\it paramagnetic} moments produced near the interface; the $H$ dependence was analyzed with Langevin or Brillouin function. The observed $\ln T$ dependence is, however, different from the Curie law ($1/T$) expected from the Langevin/Brillouin function in a low-$H$ region, which is the simplest model of localized magnetic moments \cite{takahashi, pd}. Also, similarly to the low-$H$ case, the $T$ dependence of $R_{H}$ at $9$ T is proportional to $\ln T$ in a high-$T$ regime, as shown in Fig. \ref{fig3}(c). With decreasing $T$ below $10$ K, however, $R_{H}$ at $9$ T deviates from the $\ln T$ behavior and becomes almost saturated below $2$ K, although the weak(-anti) localization does not affect the Hall effect at least in the conventional framework of weak localization. These results suggest that $\ln T$ dependence of $R_{H}$ is observed in a low-field limit, {\it i.e.} $\mu_{B}B/k_{B}T \ll 1$; since $T = \mu_{B}B/k_{B} \approx 6$ K for $B=9$ T, $R_{H}$ measured at $9$ T deviates from the $\ln T$ dependence in the low-$T$ range below $\sim 10$ K.   
\par
 
At last, anisotropy of MR is shown at $2$ K for Pt$\mid$YIG and Pt$\mid$GGG in Figs. \ref{fig4}(a) and (b), respectively, where $H$ is applied in three different directions for each sample: $H||x$, $H||y$, and $H||z$ [see also Fig. \ref{fig1}(a)]. As shown in Fig. \ref{fig4}(b), in Pt$\mid$GGG, MR, {\it i.e.} WAL, clearly depends on the $H$ direction; $|\Delta R/R|$ in $H||z$ is larger than that in in-plane $H$ cases ($H||x$ and $H||y$), which is the behavior expected from WAL in nearly two-dimensional electron systems \cite{maekawa, hoffmann, kawaguchi}. In contrast, in Pt$\mid$YIG, except for SMR contribution affected by magnetization direction in YIG in a low-$H$ region \cite{nakayama}, the high-$H$ behavior is almost isotropic with respect to $H$, as shown in symbols in Fig. \ref{fig4}(a). Since anisotropic MR is not observed even at $2$ K in our Pt$\mid$YIG, the possibility of AMR due to proximity-induced ferromagnetism in Pt \cite{huang} is ruled out. Isotropic WL observed in Pt$\mid$YIG indicates that three dimensional nature is prominent compared with Pt$\mid$GGG owing to the stronger inelastic scattering (the larger $B_{\phi}$ value) in Pt$\mid$YIG than Pt$\mid$GGG, since the condition for two dimensionality with respect to WL is that film thickness is much smaller than the dephasing length, $\sqrt{\hbar/(4eB_{\phi})}$ \cite{bergman}.  
\par

In summary, we have shown unconventional magnetotransport properties which are prominent at low temperatures and at high magnetic-fields for $1.8$-nm-thick Pt films in contact with YIG. $T$ dependence and $H$ dependence of Hall resistance are clearly affected by the interface, but not associated with those of YIG magnetization; in fact, Hall resistance shows logarithmic $T$-dependence in a broad $T$-range and nonlinear $H$ dependence at low temperatures. Also, magnetoresistance is influenced by the interface at low temperatures where quantum corrections are important, and WL behavior is observed despite the strong spin-orbit interaction of Pt. Such unconventional characteristics were not observed in Pt$\mid$GGG, although the magnitude of field-induced magnetization for GGG is comparable to that for YIG at $2$ K. 
\par

We thank Y. Fujikawa, T. Niizeki, R. Iguchi, and K. Takamura for fruitful discussions. This work was supported by CREST-JST ``Creation of Nanosystems with Novel Functions through Process Integration", PRESTO-JST ``Phase Interfaces for Highly Efficient Energy Utilization", Grants-in-Aid for JSPS Fellows, Young Scientists (B) (26790037 and 26790038), Young Scientists (A) (25707029), and Scientific Research (A) (24244051) from MEXT, and the Murata Science Foundation. Part of high magnetic-field experiments were supported by High Field Laboratory for Superconducting Materials, IMR, Tohoku University.


\begin{thebibliography}{99}
\bibitem{spin-current}S. Maekawa, S.O. Valenzuela, E. Saitoh, and T. Kimura, {\it Spin Current}, (Oxford University Press, 2012).
\bibitem{kajiwara}Y. Kajiwara, K. Harii, S. Takahashi, J. Ohe, K. Uchida, M. Mizuguchi, H. Umezawa, H. Kawai, K. Ando, K. Takanashi, S. Maekawa, and E. Saitoh, Nature(London) {\bf 464}, 262 (2010).
\bibitem{uchida}K. Uchida, J. Xiao, H. Adachi, J. Ohe, S. Takahashi, J. Ieda, T. Ota, Y. Kajiwara, H. Umezawa, H. Kawai, G.E.W. Bauer, S. Maekawa, and E. Saitoh, Nature Mater. {\bf 9}, 894 (2010).
\bibitem{huang}S.Y. Huang, X. Fan, D. Qu, Y.P. Chen, W.G. Wang, J. Wu, T.Y. Chen, J.Q. Xiao, and C.L. Chien, Phys. Rev. Lett. {\bf 109}, 107204 (2012).
\bibitem{nakayama}H. Nakayama, M. Althammer, Y.-T. Chen, K. Uchida, Y. Kajiwara, D. Kikuchi, T. Ohtani, S. Gepr\"ags, M. Opel, S. Takahashi, R. Gross, G.E.W. Bauer, S.T.B. Goennenwein, and E. Saitoh, Phys. Rev. Lett. {\bf 110}, 206601 (2013).
\bibitem{althammer}M. Althammer, S. Meyer, H. Nakayama, M. Schreier, S. Altmannshofer, M. Weiler, H. Huebl, S. Gepr\"ags, M. Opel, R. Gross, D. Meier, C. Klewe, T. Kuschel, J.-M. Schmalhorst, G. Reiss, L. Shen, A. Gupta, Y.-T. Chen, G.E.W. Bauer, E. Saitoh, and S.T.B. Goennenwein, Phys. Rev. {\bf B} {\bf 87}, 224401 (2013).
\bibitem{vlietstra}N. Vlietstra, J. Shan, V. Castel, B. J. van Wees, and J. Ben Youssef, Phys. Rev. {\bf B} {\bf 87}, 184421 (2013).
\bibitem{vlietstra-2}N. Vlietstra, J. Shan, V. Castel, J. Ben Youssef, G. E. W. Bauer and B. J. van Wees, Appl. Phys. Lett. {\bf 103}, 032401 (2013).
\bibitem{hahn}C. Hahn, G. de Loubens, O. Klein, M. Viret, V.V. Naletov, and J. Ben Youssef, Phys. Rev. {\bf B} {\bf 87}, 174417 (2013).
\bibitem{weiler-2}M. Weiler, M. Althammer, M. Schreier, J. Lotze, M. Pernpeintner, S. Meyer, H. Huebl, R. Gross, A. Kamra, J. Xiao, Y.-T. Chen, H. Jiao, G.E.W. Bauer, and S.T.B. Goennenwein, Phys. Rev. Lett. {\bf 111}, 176601 (2013).
\bibitem{isasa}M. Isasa, A. Bedoya-Pinto, F. Golmar, F. Sanchez, L.E. Hueso, J. Fontcuberta, F. Casanova, arXiv:1307.1267 (2013).
\bibitem{yang}Y. Yang, B. Wu, K. Yao, S. Shannigrahi, B. Zong, Y. Wu, arXiv:1311.1262 (2013).
\bibitem{geprags}S. Gepr\"ags, S. Meyer, S. Altmannshofer, M. Opel, F. Wilhelm, A. Rogalev, R. Gross, and S.T.B. Goennenwein, Appl. Phys. Lett. {\bf 101}, 262407 (2012).
\bibitem{qiu}Z. Qiu, K. Ando, K. Uchida, Y. Kajiwara, R. Takahashi, H. Nakayama, T. An, Y. Fujikawa and E. Saitoh, Appl. Phys. Lett. {\bf 103}, 092404 (2013).
\bibitem{bergman}G. Bergmann, Physics Reports {\bf 107}, 1-58 (1984).
\bibitem{niimi}Y. Niimi, D. Wei, H. Idzuchi, T. Wakamura, T. Kato, and Y. Otani, Phys. Rev. Lett. {\bf 110}, 016805 (2013).
\bibitem{ziman}J.M. Ziman, {\it Electrons and Phonons}, (Oxford University press, 2001)
\bibitem{hikami}S. Hikami, A.I. Larkin, and Y. Nagaoka, Prog. Theor. Phys. {\bf 63}, 707 (1980).
\bibitem{maekawa}S. Maekawa and H. Fukuyama, J. Phys. Soc. Jpn. {\bf 50}, 2516-2524 (1981).
\bibitem{dugaev}V.K. Dugaev, P. Bruno, and J. Barn\'as, Phys. Rev. {\bf B} {\bf 64}, 144423 (2001).
\bibitem{sil}S. Sil, P. Entel, G. Dumpich, and M. Brands, Phys. Rev. {\bf B} {\bf 72}, 174401 (2005).
\bibitem{kurzweil}N. Kurzweil, E. Kogan, and A. Frydman, Phys. Rev. {\bf B} {\bf 82}, 235104 (2010).
\bibitem{takahashi}S. Shimizu, K.S. Takahashi, T. Hatano, M. Kawasaki, Y. Tokura, and Y. Iwasa, Phys. Rev. Lett. {\bf 111}, 216803 (2013).
\bibitem{pd}T. Lin, C. Tang, and J. Shi, Appl. Phys. Lett. {\bf 103}, 132407 (2013).
\bibitem{ahe-review}N. Nagaosa, J. Sinova, S. Onoda, A.H. MacDonald, and N.P. Ong, Rev. Mod. Phys. {\bf 82}, 1539 (2010).
\bibitem{the}P. Bruno, V.K. Dugaev, and M. Taillefumier, Phys. Rev. Lett. {\bf 93}, 096806 (2004).
\bibitem{hoffmann}H. Hoffmann, F. Hofmann, and W. Schoepe, Phys. Rev. {\bf B} 25, 5563 (1982).
\bibitem{kawaguchi}T. Kawaguchi and Y. Fujimori, J. Phys. Soc. Jpn. {\bf 52}, 722-725 (1983).
\end{thebibliography}
\end{document}